# Exploring the Use of Contingency Analysis for Nuclear Electrical Studies

Cameron Khanpour, Jon T. Fontejon
*School of Electrical and Computer Engineering*
*Georgia Institute of Technology*

*Abstract*—This paper examines the use of contingency analysis for a nuclear power plant to determine its potential benefits for the nuclear industry. Various N-1 contingencies were analyzed for a model of an existing nuclear plant, primarily inspecting voltage violations resulting from a failure. Remedial Actions Schemes were suggested to support the reduction of voltage violations in the event of a failure within the system. Many of the schemes presented were solved by existing redundancies and protection schemes that have been provided through the use of industry standard bounding analysis in the design process. This paper proposes the future use of real-time contingency analysis for nuclear power plants, conducted using constantly updating voltage, current, and power measurements through the system. This will provide real-time information of the system and can serve as historical data to reduce the analysis needed for pending design changes in the plant.

## I. Introduction

Nuclear power plants stand unique in the vast landscape of power generation, presenting unparalleled operational intricacies and challenges. Central to the functioning of these nuclear plants is the inherent need for a constant power supply. Unlike other generation sources where a trip may lead to simple shutdown procedures, the cessation of a nuclear plant presents a more complex challenge. It's imperative to continuously power the coolant pumps to keep the reactor core cooled until it's self-sufficient through natural circulation.

However, despite these evident demands, the nuclear industry has traditionally steered away from real-time contingency analysis. This can be attributed to the inbuilt redundant power trains, which, from a regulatory standpoint, often meet the licensing benchmarks such as the Technical Specifications (Tech Specs). Furthermore, the worst-case scenarios, which dictate plant configurations for events like outages or accident conditions such as a LOCA (loss of coolant accident), tend to act as safety cushions. While these established procedures provide strong electrical security, there remains an unexplored potential in real-time contingency analysis. This could reshape how the industry approaches both plant safety and operational efficiencies.

In this paper, we will analyze an example model of the electrical distribution of a nuclear power plant. We will utilize the analysis software Electrical Transient Analyzer Program (ETAP) to run a python script that will execute a number of load flow studies covering multiple potential contingencies. Then we will show the results of voltage violations beyond ±10% nominal voltage for a given contingency and describe potential Remedial Action Schemes (RAS).

## II. Background

### A. Basic Definitions

An illustration of a modeled typical one-line diagram for the operations of a single unit in a nuclear power system can be seen in Fig. 1. The left side of the one line starting with "XFMR1" handles the internal operations of the system, including the various switchgears (SWGR) and motor control centers (MCC). The right side of the one line starting with "Step-up XFMR" handles the external transmission of power, starting at the system switchyard and sending it through transmission lines beyond the plant. This paper will discuss contingencies on both sides of the one-line.

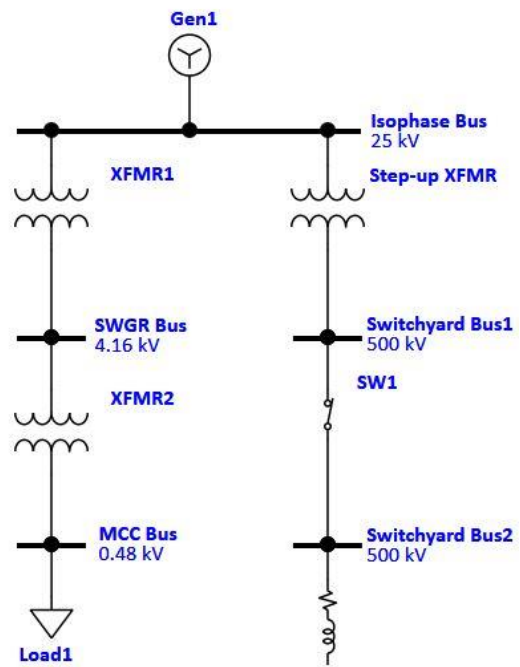

Fig. 1. Typical one-line diagram for a nuclear system.

## B. Literature Review

As one of the few papers in the IEEE database discussing contingency analysis on nuclear power systems, [1] serves as a strong basis for the methods of this paper. Its utilization of ETAP and N-1 contingency analysis makes it a valuable resource. However, it only considers three N-1 line contingencies, which limits its analysis capabilities. Furthermore, all of the analyzed contingencies meet standard grid code criteria and operating limits, requiring no corrective action within the system or any suggestions from the authors to enhance security.

Similar papers that discuss the stability of nuclear power systems are [4] and [8], though provide more background information than results to be used. The former delves into faults specific to nuclear plants, such as RCP rotor stuck accident, Loss of Feedwater Accident, and Loss of Coolant Accident. For the purposes of this paper, these faults could be considered an N-1 generator contingency. The latter discusses the importance of stable offsite power to the functions of nuclear power plants. Safety analyses typically assume the electric grid is stable, allowing plants to follow their safety procedures and corrective actions. The greatest weakness of offsite power are the thermal limits of existing transmission lines to nuclear plants, which can be modeled as N-1 line contingencies.

Aside from these three resources, there are few relevant papers that discuss contingency analysis specifically for nuclear systems. Therefore, it will be beneficial to look at contingency analysis for general power systems, for which there is a multitude of documentation. Works such as [3], [9], and [10] focus on methods to conduct contingency analyses, which make them valuable resources for the methodology of this paper. In particular, [3] and [10] also use ETAP, where the former analyzes seven different contingencies, including a normal operation case for comparison, and the latter analyzes an IEEE 14-bus system for ten N-1 contingencies on transmission line failures.

Works such as [2], [5], [6], and [7] are assessments that utilize contingency analysis for their conclusions. While this paper is not an assessment of any specific system, the use of contingency analysis makes these papers worth noting. The most relevant of these works are [2], which uses ETAP to eleven different generator contingencies to analyze a 470-bus system of varying load, and [7], which conducts N-1 line contingency analysis on 80 transmission lines and N-1 contingency analysis on seven generators in an IEEE 57-bus system. The former uses contingency analysis to evaluate the frequency response of the system under the different contingencies and the latter ranks each contingency by the extent to which it overloads the system. While this paper will not be looking at the system frequency response, it will also rank each contingency by its overvoltage results.

Case studies that discuss processes either before or after contingency analysis could be valuable for further investigation into contingency analysis for nuclear power systems. Equivalent networks can be developed using Ward and REI reduction methods before analyzing contingencies [6]. This paper will not be utilizing any reduction methods, but it may be worth investigating to speed up the contingency analysis process. To obtain a more comprehensive evaluation, N-1-1 contingency analysis can be conducted, where a second outage occurs after corrective action is taken to remedy a N-1 contingency [5]. While this paper will not be investigating N-1-1 contingencies, its conclusions could support further analyses into cases where corrective action is needed.

## III. PROPOSED METHOD

### A. Mathematical Model and Formulation

ETAP allows the option to conduct load flow analyses using Gauss-Seidel, Newton-Raphson, or Fast Decoupled. The default option is Newton-Raphson (NR), which is the method that will be used in this paper. This method solves the following equations iteratively,

$$P_i = \sum_{j=1}^{n} |Y_{ij}||V_i||V_j| \cos(\theta_{ij} + \delta_i + \delta_j) \quad (1)$$

$$Q_i = -\sum_{j=1}^{n} |Y_{ij}||V_i||V_j| \sin(\theta_{ij} + \delta_i + \delta_j) \quad (2)$$

When equations 1 and 2 are expanded using Taylor's series about an initial estimate and neglecting higher order terms, the following set of linear equations is obtained,

$$\begin{bmatrix} \Delta P_2^{(k)} \\ \vdots \\ \Delta P_n^{(k)} \\ \Delta Q_2^{(k)} \\ \vdots \\ \Delta Q_n^{(k)} \end{bmatrix} = \begin{bmatrix} \frac{\partial P_2^{(k)}}{\partial \delta_2} & \cdots & \frac{\partial P_2^{(k)}}{\partial \delta_n} & \frac{\partial P_2^{(k)}}{\partial |V_2|} & \cdots & \frac{\partial P_2^{(k)}}{\partial |V_n|} \\ \vdots & \ddots & \vdots & \vdots & \ddots & \vdots \\ \frac{\partial P_n^{(k)}}{\partial \delta_2} & \cdots & \frac{\partial P_n^{(k)}}{\partial \delta_n} & \frac{\partial P_n^{(k)}}{\partial |V_2|} & \cdots & \frac{\partial P_n^{(k)}}{\partial |V_n|} \\ \frac{\partial Q_2^{(k)}}{\partial \delta_2} & \cdots & \frac{\partial Q_2^{(k)}}{\partial \delta_n} & \frac{\partial Q_2^{(k)}}{\partial |V_2|} & \cdots & \frac{\partial Q_2^{(k)}}{\partial |V_n|} \\ \vdots & \ddots & \vdots & \vdots & \ddots & \vdots \\ \frac{\partial Q_n^{(k)}}{\partial \delta_2} & \cdots & \frac{\partial Q_n^{(k)}}{\partial \delta_n} & \frac{\partial Q_n^{(k)}}{\partial |V_2|} & \cdots & \frac{\partial Q_n^{(k)}}{\partial |V_n|} \end{bmatrix} \begin{bmatrix} \Delta \delta_2^{(k)} \\ \vdots \\ \Delta \delta_n^{(k)} \\ \Delta |V_2^{(k)}| \\ \vdots \\ \Delta |V_n^{(k)}| \end{bmatrix} \quad (3)$$

Or in short form,

$$\begin{bmatrix} \Delta P \\ \Delta Q \end{bmatrix} = \begin{bmatrix} J_1 & J_2 \\ J_3 & J_4 \end{bmatrix} \begin{bmatrix} \Delta \delta \\ \Delta |V| \end{bmatrix}$$

Where $\Delta P$ and $\Delta Q$ represent the difference between specified and calculated values of real and reactive powers respectively. $J_1, J_2, J_3$, and $J_4$ are the Jacobian matrix elements. $\Delta \delta$ and $\Delta |V|$ are the bus angle vectors and voltage magnitudes. These values are given an initial estimate and will be updated with every iteration until it is stable within a specified limit. After the solution of the above-mentioned equations, the results of load flow analysis are obtained.

In addition, in the load flow analysis, contingencies are placed on the model, such as failure of transformers, buses, or breakers. With one of these contingencies in place, load flow analysis is performed to display what voltage violations are present in the model. Voltage violations are typically determined by the plants licensing; however, we are reporting ±10% of nominal voltage for over and undervoltage respectively.

### B. Proposed Solution Method

Two integral components are needed to conduct contingency analysis: a model and a list of contingencies. For the sake of this experiment, only one model will be used. This model will be similar to a typical Pressurized Water Reactor (PWR) electrical system. A small section of this model, with steady-state load flow, can be seen in Figure 2.

Contingencies are also needed for analysis purposes. These contingencies primarily involve the failure of components within the model; however, hypothetical contingencies that involve systems outside of the scope of the model are also considered. The list of contingencies to be analyzed can be seen in Table 1.

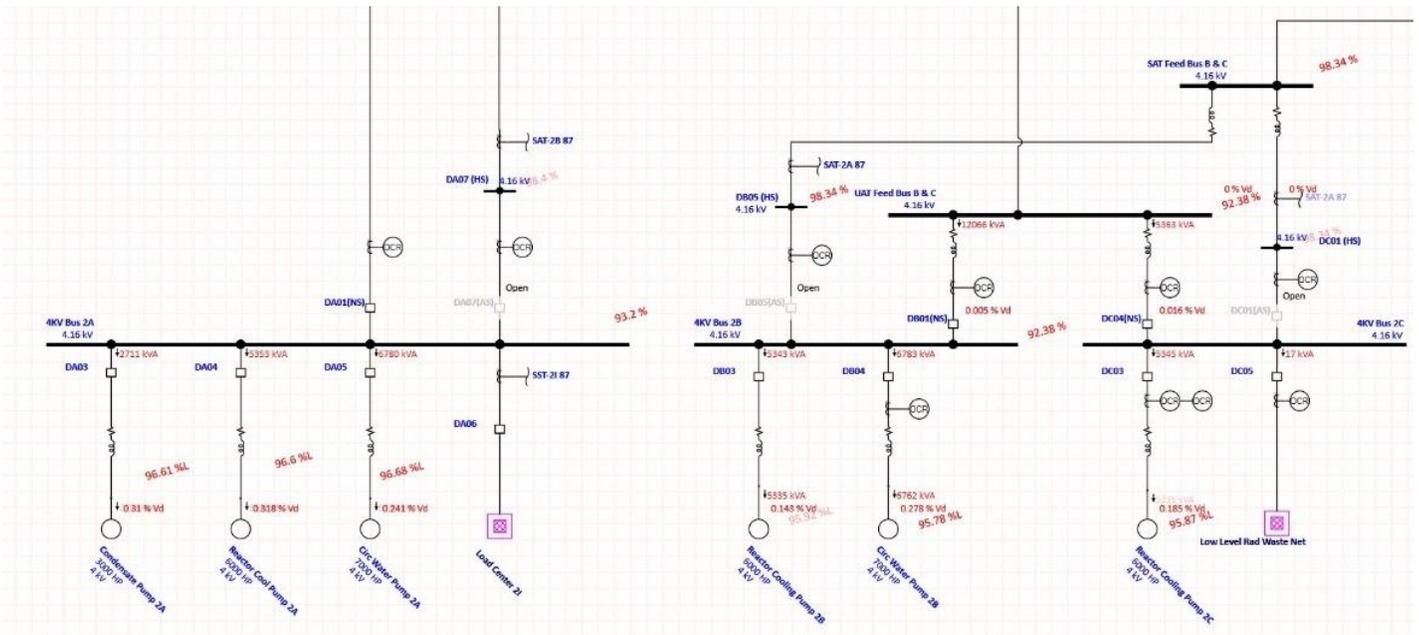

Fig. 3. Section of 4kV level of electrical distribution system of a typical PWR under stead state load flow.

| Contingency | Type | Description of Failure |
|---|---|---|
| SAT 2B Y Winding Failure | N-1 | Transformer Winding |
| SAT 2B Y Winding turned on during LOCA | N-1 | Transformer Winding Unintentionally On |
| Change in System Impedance During Outage | N-1 | Decrease in System Impedance |
| Transmission Line Failure | N-1 | Hypothetical Failure of Transmission Line |
| Tie Breaker Failure | N-1 | Hypothetical Failure of Tie Breaker |
| Incoming SAT Breaker Failure Transferring to UAT | N-1 | Breaker Failure |

Table 1. Contingencies used for analysis.

The method to conduct contingency analysis will involve imposing a contingency failure on the model before running load-flow analysis on the changed model. Voltage violations will be recorded before resetting the model to its pre-contingency state.

IV. SIMULATION RESULTS

A. Case Description

**SAT 2B Y Winding Failure:**
During normal operation, the SAT 2B Y Winding powers downstream 4.16kV buses 2E and 2G. This contingency looks at the failure of this winding during normal operation.

Results in a failure of 4kV buses 2E and 2G and a voltage violation at each of their respective 600V and 480V load centers.

The Remedial Action Scheme (RAS) is to perform a Fast Bus Transfer from SAT 2B to UAT 2B. This results in a slight voltage drop (~1%) from the steady state load flow, however it is still within the allowable margin for operation.

**SAT 2B Y Winding turned on during LOCA:**
This is a Loss-of-Coolant Accident (LOCA) condition in which motors are turned on to reduce heat within the plant, such as residual heat removers (RHR pumps), causing an increase in load on the system. Typically, within a plant, there are two sets of cooling fans in each unit for redundancy. During this contingency, we are assuming that the SAT 2B Y winding was left on due to a failure to swing the loads to the other unit, increasing the load on the transformer.

This results in undervoltage violations on the loads on the downstream 4kV 2E and 2G buses, which includes 600V Load Centers 2W, 2U, 2Z, 2X, 2V, 2E, 2D, and 600V MCCs 1G, 2T, 2B, 2DD, 2V, 1F, 2CC. The loads on these load centers mainly consist of cooling tower fans.

The RAS is to manually open breakers DE03 and DG15 to shed loading. The cooling tower fans on 4kV 2D and 2F buses can be used to continue to reduce heat in the plant. This removes the voltage violations from the load flow analysis.

**Change in System Impedance:**
During normal analysis conditions, system impedance is assumed to be a constant, non-zero value. However, during an outage, this system impedance is subject to change, with the worst-case scenario being a system impedance of zero.

A system impedance of zero results in overvoltage violations on 208V MCC 2M, 1N, 2A, 2B, 2D, 2E. These loads consisted of radiation monitors, space heaters, and fans. These are non-safety related loads, meaning that they are not vital for the shutdown of the plant.

The RAS is to apply temporary power from another switchgear to power these loads. This removes the voltage violations from the load flow analysis.

**Transmission Line Failure:**

There is a transmission line connected to the plant with a capacity of 2500 MW, and the modeled plant generates 1800 MW. Suppose under a hypothetical N-1 contingency, the line capacity decreases to 1200 MW. To match this constraint, the nuclear power plant must decrease its production to prevent overloading the line. However, nuclear power plants are used to power critical infrastructure such as hospitals due to the plant's consistent power output. As such, if the nuclear power plant decreases its production below a set value, these critical infrastructures will lose the ability to operate.

Therefore, the RAS is to contact grid operators to increase production in other plants such as gas power plants to both compensate for loss of transmission capacity and sustain operation of critical infrastructures.

**Tie Breaker Failure:**

Between a safety-related 600V bus and a non-safety related 600V bus, there is a tie breaker connecting them. They are usually connected by a normal source 4.16kV bus, with an alternate source from a 600V bus. In the event of an accident failure, Loss of Offsite Power, where the grid goes offline, the loads would be powered by diesel generators. During this time, the tie breaker would open and isolate the non-safety bus to be powered by the 4.16kV bus, and the safety-related bus would be powered by the alternate 600V bus.

In the event the tie breaker fails and stays closed, this would result in a dual-source-powered bus, causing current to flow back through the 600V bus into the rest of the system, leading to overvoltage violations throughout the system.

The RAS is to open the breaker of the normal 4.16kV bus, ensuring there is only one source powering the safety-related bus.

**Incoming SAT Breaker Failure:**

4.16kV buses have two breakers connected to them to represent two different sources connecting to the bus, which are the start-up auxiliary transformers (SAT) and unit auxiliary transformers (UAT). During normal operation, the buses are sourced from the SATs.

During a failure of the generator on the SAT side, the incoming breaker connecting the SATs to the 4.16kV bus is opened, and the breaker connecting the UATs to the 4.16kV bus is closed. This is transferred via a supervised fast bus transfer scheme.

In the event of an incoming breaker failure where both the incoming SAT breaker and the UAT breaker are both open, there will be an open circuit in the system, resulting in undervoltage violations in all downstream components. The RAS would be to conduct a supervised fast bus transfer to the UATs.

*B. Numerical Results*

The buses analyzed for voltage violations are vital load center and MCC buses. Many of the buses in the analysis reports are redundant nodes between components in the model.

Relevant sections of contingency analysis results are provided in Table 2 and Table 3 below. Overvoltage and undervoltage violations of $\pm 10\%$ of the nominal bus voltage are seen as significant. Table 2 shows the extent of undervoltage violations for the SAT 2B Y Winding turned on during LOCA contingency, providing the percentage of the nominal voltage of each bus due to failure.

| | Bus ID | Nominal kV | Voltage |
|---|---|---|---|
| 109 | 600V Load Center 2W | 0.6 | 88.6 |
| 110 | Bus337 | 0.6 | 88.6 |
| 111 | Bus415 | 0.6 | 88.62 |
| 112 | Bus673 | 0.6 | 88.62 |
| 113 | Bus488 | 0.6 | 88.64 |
| 114 | Bus513 | 0.6 | 88.64 |
| 115 | Bus724 | 0.6 | 88.64 |
| 116 | Compskid Mtr B-Trn (CU2) | 0.6 | 88.64 |
| 117 | HS 600/208V Xfmr MCC 1G | 0.6 | 88.67 |
| 118 | Bus67 | 0.6 | 88.68 |
| 119 | Bus131 | 0.6 | 88.68 |
| 120 | 600V MCC 1G | 0.6 | 88.7 |
| 121 | Bus342 | 0.6 | 88.73 |
| 122 | Bus169 | 0.6 | 88.74 |
| 123 | Bus191 | 0.6 | 88.75 |
| 124 | 600V Load Center 2U | 0.6 | 88.76 |
| 125 | Bus346 | 0.6 | 88.77 |
| 126 | Bus439 | 0.6 | 88.83 |
| 127 | Bus99 | 0.6 | 88.84 |
| 128 | HS Dies 2B Aux Trans | 0.6 | 88.9 |
| 129 | 600V MCC 2T | 0.6 | 88.92 |
| 130 | 600V Load Center 2Z | 0.6 | 88.93 |

Table 2. Voltage violations due to SAT 2B Y Winding turned on during LOCA.

Table 3 shows the undervoltage violations after the RAS has been conducted in the system for the SAT 2B Y Winding contingency. The RAS has partially remedied the undervoltage violations, allowing buses to return within the $\pm 10\%$ criteria.

| Bus | Nominal kV | Voltage |
|---|---|---|
| 208V MCC 2AA | 0.208 | 91.32 |
| 208V MCC 2BB | 0.208 | 91.73 |
| 208V MCC 2W | 0.208 | 91.6 |
| 208V Swgr 2A | 0.208 | 91.78 |
| 208VTurb Bldg Dist Cab 2J | 0.208 | 91.45 |
| 208VTurb Bldg DistCab 2GG | 0.208 | 91.56 |
| 208VTurb Bldg DistCab 2H* | 0.208 | 91.5 |
| 600V Load Center 1K | 0.6 | 91.49 |
| 600V Load Center 1L | 0.6 | 90.28 |
| 600V Load Center 2A | 0.6 | 91.34 |
| 600V Load Center 2B | 0.6 | 91.62 |
| 600V Load Center 2C | 0.6 | 90.23 |
| 600V Load Center 2E Emerg | 0.6 | 90.59 |
| 600V Load Center 2K | 0.6 | 91.49 |
| 600V Load Center 2L | 0.6 | 90.28 |
| 600V Load Center 2Q | 0.6 | 91.32 |

Table 3. Voltage violations after remedial action for SAT 2B Y Winding Failure.

*C. Discussion*

The implications of the results show that considering specific contingencies can be beneficial in order to provide Remedial

5Action Schemes to operators or to help design automatic functions to handle specific failures. These results can also be used to support design changes to optimize the network for further system reliability.

Some limitations of using contingency analysis in the lens of the nuclear industry is that some of the problems that contingency analysis can solve are already solved from redundancies and worst-case bounding analysis. However, we believe this can help lay some of the groundwork towards utilizing contingency analysis to help replace bounding analysis in favor of realistic failures. Some possible future work to help reach that goal is incorporate Probabilistic Risk Analysis (PRA) in the electrical distribution system to determine the most probable and/or most consequential events and use that as a basis for the contingencies.

Additionally, the intent of this work is to propose the use of real-time contingency analysis for nuclear plants. This would be done by constantly capturing voltage current, and power measurements throughout the electrical distribution of the plant. The data would be sent to a monitoring and diagnostics center where real-time contingency analysis would be performed against the updating data over a list of known contingencies that would adapt to the condition of the plant. This differs from how plant configuration is analyzed in the nuclear industry, where only certain configurations are analyzed as a bounding case, such as LOCA or Outage configurations. Real time contingency analysis and power consumption data can also be used as a reference for a certain amount of time, such as the past two years for outage cycles, acting as historical data to determine whether adding or removing load from the system would impact any voltage violations of other loads. This would significantly reduce analysis needed for pending design changes in the plant.

## V. Conclusions

A brief analysis of different contingencies for a nuclear power plant has been presented. Six N-1 contingencies were considered, including two hypothetical contingencies due to the limitations of the ETAP model. The Remedial Action Schemes for these contingencies were shown to reduce voltage violations in the system. The analyzed contingencies were largely solved by existing protection schemes and worst-case bounding analysis. These procedures make nuclear plants more robust than other generation plants and remove the need for nuclear plants to conduct contingency analysis. However, the results of this work show that considering specific contingencies provides Remedial Action Schemes for nuclear plants and could accelerate a return to normal operation for a plant in the event of a failure. The use of real-time contingency analysis for nuclear plants has been proposed, conducted using constantly updating voltage, current, and power measurements throughout the system. This analysis differs from the current industry standard, which only examines certain configurations as a bounding, worst-case incident, and does not provide real-time information in the event of a failure. The data generated from real-time contingency analysis can act as historical data to determine whether adding or removing load from the system would impact any voltage violations of other loads, reducing the analysis needed for pending design changes in the plant. The additional reliability provided from contingency analysis can serve as a powerful motivating factor for the nuclear industry to shift from bounding analysis to real-time analysis.

## Acknowledgments

This paper was presented at ETAP Nuclear Utility Users Conference (NUUC) at Irvine, California, June 18th, 2024. The authors would like to thank the conference organizers and attendees for their valuable feedback.
## References

[1] W. Irfan, M. Awais, D. N. Zareen and I. Ahmed, "N-1 Contingency Analysis for Offsite Power System of an HPR-1000 Power Plant Using ETAP Software," 2022 International Conference on Recent Advances in Electrical Engineering & Computer Sciences (RAEE & CS), Islamabad, Pakistan, 2022, pp. 1-5, doi: 10.1109/RAEECS56511.2022.9954560.

[2] Liang Wang, Li Li, Shanshan Shi, Yiwei Zhang, Zongxiang Lu and Junliang Zhang, "Stability and security assessment for an industrial electric grid with enterprise-owned power plants," 2008 Third International Conference on Electric Utility Deregulation and Restructuring and Power Technologies, Nanjing, 2008, pp. 1559-1563, doi: 10.1109/DRPT.2008.4523653.

[3] R. A. Gamboa, C. V. Aravind and C. A. Chin, "Power System Network Contingency Studies," 2018 IEEE Student Conference on Research and Development (SCOReD), Selangor, Malaysia, 2018, pp. 1-6, doi: 10.1109/SCORED.2018.8711362.

[4] Q. Shi, D. Liu, H. Luo and Y. Gao, "Study on the faults of nuclear power plant and effects on transient stability of power system," 2012 China International Conference on Electricity Distribution, Shanghai, China, 2012, pp. 1-5, doi: 10.1109/CICED.2012.6508682.

[5] P. S. V. Prabhakar, R. Krishan and D. R. Pullaguram, "Static Security Assessment of Large Power Systems Under N-1-1 Contingency," 2022 22nd National Power Systems Conference (NPSC), New Delhi, India, 2022, pp. 35-40, doi: 10.1109/NPSC57038.2022.10069705.

[6] M. Ramirez-Gonzalez, M. Bossio, F. R. S. Sevilla and P. Korba, "Evaluation of Static Network Equivalent Models for N-1 Line Contingency Analysis," 2022 4th Global Power, Energy and Communication Conference (GPECOM), Nevsehir, Turkey, 2022, pp. 328-333, doi: 10.1109/GPECOM55404.2022.9815713.

[7] K. A. Reddy Medapati, S. Mandal, R. Paul, A. Samanta, D. Bose and A. Chakrabarti, "Assessment of Power System Security using Contingency Ranking Analysis," 2023 5th International Conference on Energy, Power and Environment: Towards Flexible Green Energy Technologies (ICEPE), Shillong, India, 2023, pp. 1-6, doi: 10.1109/ICEPE57949.2023.10201584.

[8] N. K. Trehan, "Electrical grid stability and its impact on nuclear power generating stations," IECEC-97 Proceedings of the Thirty-Second Intersociety Energy Conversion Engineering Conference (Cat. No.97CH6203), Honolulu, HI, USA, 1997, pp. 1744-1747 vol.3, doi: 10.1109/IECEC.1997.656685.

[9] S. Nisworo, S. Setiawan, D. Pravitasari, A. Trihasto and Z. A. Kusworo, "Contingency Analysis of Electric Power Flow as a Power System Operation Approach," 2022 5th International Conference on Power Engineering and Renewable Energy (ICPERE), Bandung, Indonesia, 2022, pp. 1-5, doi: 10.1109/ICPERE56870.2022.10037422.

[10] A. E. Airoboman, P. James, I. A. Araga, C. L. Wamdeo and I. K. Okakwu, "Contingency Analysis on the Nigerian Power Systems Network," 2019 IEEE PES/IAS PowerAfrica, Abuja, Nigeria, 2019, pp. 70-75, doi: 10.1109/PowerAfrica.2019.8928883.
5